\documentclass[utf8]{article}
\usepackage{import}
\pdfoutput=1 

\usepackage[english]{babel}
\usepackage[T1]{fontenc}

\usepackage[a4paper, top=3cm,bottom=2cm,left=3cm,right=3cm,marginparwidth=4cm]{geometry}
\usepackage{stringstrings}
\usepackage[explicit]{titlesec}
\usepackage[dvipsnames,table,xcdraw]{xcolor}
\usepackage{subfigure}
\usepackage[subfigure]{tocloft}
\usepackage[round, sort]{natbib}
\usepackage{nameref}
\usepackage{amsmath}
\usepackage{amsfonts}
\usepackage{graphicx}
\usepackage{svg}
\usepackage{authblk}
\usepackage{float}
\usepackage{tikz}
\usepackage{lineno}
\usepackage{csquotes}

\usepackage[colorinlistoftodos]{todonotes}
\usepackage[at]{easylist}
\usepackage{tcolorbox}
\tcbuselibrary{skins}
\usepackage{comment}
\usepackage[framemethod=tikz]{mdframed}
\usetikzlibrary{shadows}
\usepackage{soul}
\usepackage{version}

    	{
            \begin{tcolorbox}[enhanced,
                title=-,
                size=small,
                colbacktitle=Aquamarine,
                drop fuzzy shadow,
                fontupper=\small,
                boxrule=0.4pt,
                colback=Aquamarine!10!white,
                sharp corners]
                Writing Materials
            \end{tcolorbox}
            \begin{easylist}[itemize]
    	}
    	{
            \end{easylist}  
            \begin{tcolorbox}[enhanced,
                halign=flush right,
                halign title=right,
                size=small,
                colbacktitle=Aquamarine,
                drop fuzzy shadow,
                fontupper=\small,
                boxrule=0.4pt,
                colback=Aquamarine,
                colupper=White,
                sharp corners]
                Writing Materials -
            \end{tcolorbox}        
    	}	
    	
        \newcommand{\needfig}[1]{%
			\ifthenelse{\equal{#1}{}}{%
				\todo[color=White, linecolor=Orange, bordercolor=RedOrange]{\textcolor{RedOrange}{Fig}}}{%
				\todo[color=White, linecolor=Orange, bordercolor=RedOrange]{\textcolor{RedOrange}{Fig: #1}}%
			}%
		}

		\newcommand{\needref}[1]{%
			\ifthenelse{\equal{#1}{}}{%
				\todo[color=White, linecolor=Orange, bordercolor=Orange]{\textcolor{Orange}{Ref}}}{%
				\todo[color=White, linecolor=Orange, bordercolor=Orange]{\textcolor{Orange}{Ref: #1}}%
			}%
		}

    \usepackage[xcolor, outerbars]{changebar}
    \cbcolor{red}
    
\usepackage{amsthm}

\newlength{\mylen}
\renewcommand{\cftfigpresnum}{\figurename\enspace}
\renewcommand{\cftfigaftersnum}{:}
\newtheorem{definition}{Definition}
\newtheorem*{hypothesis}{Hypothesis}

\usepackage{chngcntr}
\counterwithout{figure}{section}

\usepackage{acronym}
\newacro{ntic}[NTIC]{non-trivial informational closure}
\newacro{ic}{informational closure}
\newacro{cg}{coarse-grain}
\newacro{cging}{coarse-graining}
\newacro{cged}{coarse-grained}
\newacro{ncg}{neural coarse-graining}
\newacro{OurTheory}[ICT]{Information Closure Theory of Consciousness}

\newcommand{\X}{\mathcal{X}}
\renewcommand{\S}{\mathcal{S}}
\newcommand{\E}{\mathcal{E}}

\title{Information Closure Theory of Consciousness}
\date{}

\author[]{Acer Y.C. Chang\thanks{Corresponding author, acercyc@araya.org}}
\author[]{Martin Biehl\thanks{martin@araya.org}}
\author[]{Yen Yu\thanks{yen.yu@araya.org}}
\author[]{Ryota Kanai\thanks{kanair@araya.org }}
\affil[]{ARAYA, Inc., Tokyo, Japan}

\begin{document}
%    \linenumbers
	\maketitle
	\tableofcontents

	\begin{abstract}
		Information processing in neural systems can be described and analysed at multiple spatiotemporal scales. Generally, information at lower levels is more fine-grained but can be coarse-grained at higher levels. However, only information processed at specific scales of coarse-graining appears to be available for conscious awareness. We do not have direct experience of information available at the scale of individual neurons, which is noisy and highly stochastic. Neither do we have experience of more macro-scale interactions, such as interpersonal communications. Neurophysiological evidence suggests that conscious experiences co-vary with information encoded in coarse-grained neural states such as the firing pattern of a population of neurons. In this article, we introduce a new informational theory of consciousness: \acf{OurTheory}. We hypothesise that conscious processes are processes which form \ac{ntic} with respect to the environment at certain coarse-grained scales. This hypothesis implies that conscious experience is confined due to informational closure from conscious processing to other coarse-grained scales. \ac{OurTheory} proposes new quantitative definitions of both conscious content and conscious level. With the parsimonious definitions and a hypothesise, \ac{OurTheory} provides explanations and predictions of various phenomena associated with consciousness. The implications of \ac{OurTheory} naturally reconcile issues in many existing theories of consciousness and provides explanations for many of our intuitions about consciousness. Most importantly, \ac{OurTheory} demonstrates that information can be the common language between consciousness and physical reality.

	\end{abstract}

	\section*{Keywords:}
	Keywords: theory of consciousness, non-trivial informational closure, NTIC, coarse-graining, level of analysis

% ============================================================================ %
%                                     Start                                    %
% ============================================================================ %

    \newpage
	\section{Introduction}

		%% Story about a cell
		Imagine you are a neuron in Alice’s brain. Your daily work is to collect neurotransmitters through dendrites from other neurons, accumulate membrane potential, and finally send signals to other neurons through action potentials along axons. However, you have no idea that you are one of the neurons in Alice’s supplementary motor area and are involved in many motor control processes for Alice’s actions, such as grabbing a cup. You are ignorant of intentions, goals, and motor plans that Alice has at any moment, even though you are part of the physiological substrate responsible for all these actions. A similar story also happens in Alice’s conscious mind. To grab a cup, for example, Alice is conscious of her intention and visuosensory experience of this action. However, her conscious experience does not reflect the dynamic of your membrane potential or the action potentials you send to other neurons every second. That is, not all the information you have is available to Alice’s conscious mind.

		%% scale
		
\begin{figure}[H]
		    \centering
			\includegraphics[width=\textwidth]{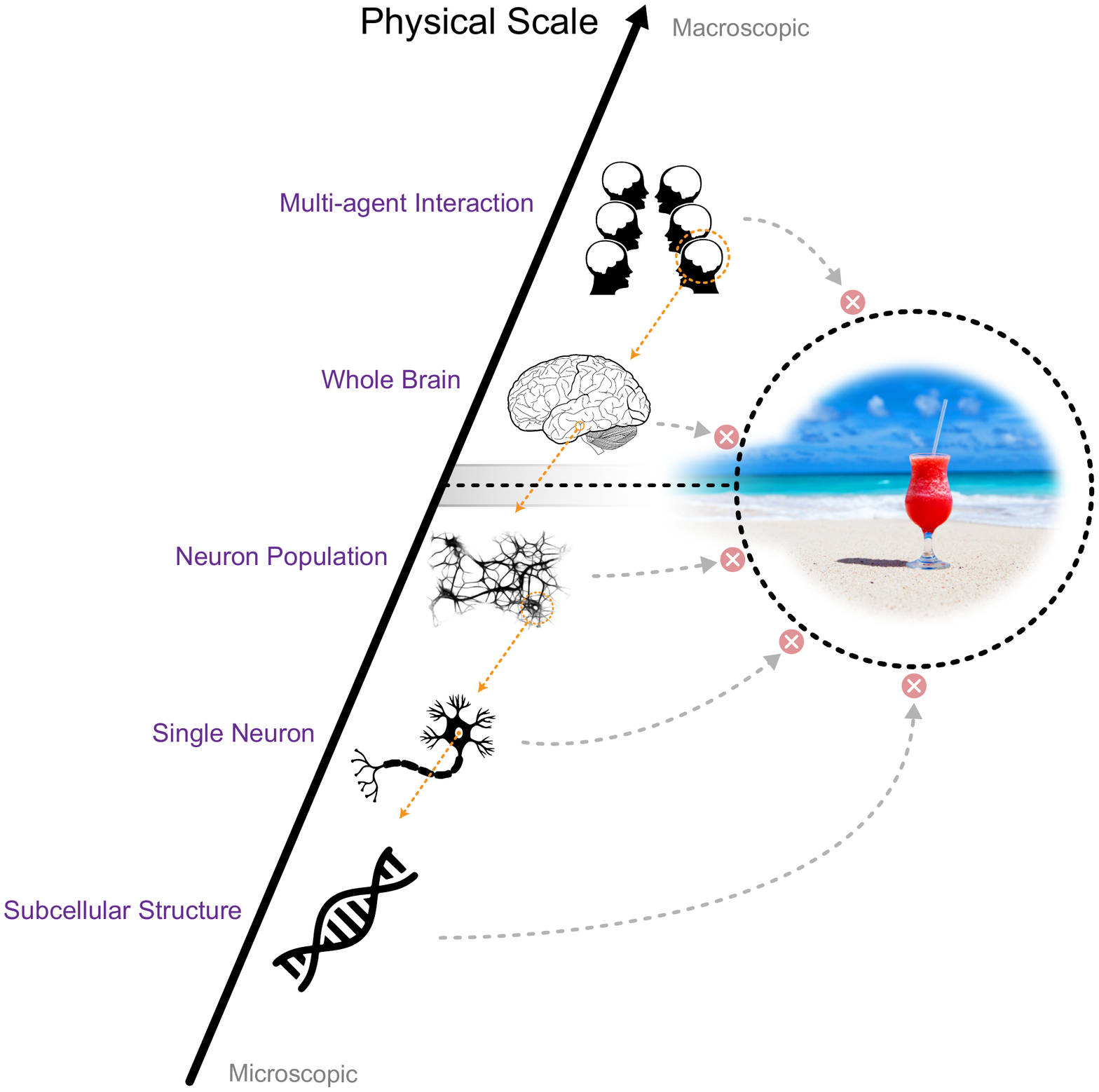}
			\caption{The scale problem of consciousness: Human conscious experience does not reflect information from every scale. Only information at a certain coarse-grained scale in the neural system is reflected in consciousness.}
			\label{fig:scaleproblem}
	   	\end{figure}

		It appears to be true that we do not consciously access information processed at every scale in the neural system. There are both more microscopic and more macroscopic scales than the scale corresponding to the conscious contents. On the one hand, the dynamics of individual neurons are stochastic \citep{Goldwyn2011, White2000}. However, what we are aware of in our conscious mind shows astonishing stability and robustness against the ubiquitous noise in the neural system \citep{mathis1995computational}. In addition, some parts of the neural system contribute very little to conscious experience (the cerebellum for example \citep{lemon2010life}), also suggesting that conscious contents do not have one-to-one mapping to the entire state of the neural system. On the other hand, human conscious experience is more detailed than just a simple (e.g. binary) process can represent, suggesting that the state space of conscious experience is much larger than what a single overly coarse-grained binary variable can represent. These facts suggest that conscious processes occur at a particular scale. We currently have possess only a few theories (e.g., Integrated Information Theory \citep{hoel2016can} and Geometric Theory of Consciousness \citep{fekete2011towards,fekete2012lack}) to identify the scale to which conscious processes correspond (also see discussion in \cite{fekete2016system}). We refer to this notion as \textbf{the scale problem of consciousness} (Fig.~\ref{fig:scaleproblem}).

		%% our argument
		In this article, we propose a new information-based theory of consciousness, called the  \acf{OurTheory}. We argue that every process with a positive non-trivial information closure (NTIC) has consciousness. This means that the state of such a process corresponds one-to-one to conscious content.\footnote{In the following IC stands for "informational closure" or "informationally closed" and NTIC stands for "non-trivial informational closure" or "non-trivially informationally closed".}. We further postulate that the \textit{level} of consciousness corresponds to the degree of NTIC. (For a discussion of the distinction between level versus content of consciousness see \cite{laureys2005neural, overgaard2010neural}).
		
		In the following, we first introduce non-trivial informational closure and argue for its importance to information processing for human scale agents (Sec.~\ref{sec:Non-trivial informational closure}). We next argue that through coarse-graining the neural system can form informational closure and a high degree of NTIC at a specific scale of coarse-graining  (Sec.~\ref{sec:Neural coarse-graining}). In Sec.~\ref{sec:OurTheory}, we propose a new theory of consciousness (\ac{OurTheory}). We also illustrate how \ac{OurTheory} can parsimoniously explain empirical findings from previous consciousness studies (Sec.~\ref{sec:Conscious versus Unconscious Processing}) and reconcile several current major theories of consciousness (Sec.\ref{sec:Comparison with other theories}). Finally, we discuss the current theoretical and empirical limitations of \ac{OurTheory} and propose the implications of \ac{OurTheory} on the current consciousness science (Sec.\ref{sec:Limitation and Future work}).

% ============================================================================ %
%                       Non-trivial informational closure                      %
% ============================================================================ %
	\section{Non-trivial Informational Closure} \label{sec:Non-trivial informational closure}
		The notion of non-trivial informational closure (NTIC) was introduced by \cite{BERTSCHINGER.2006}. The concept of closure is closely related to system identification in systems theory. One can distinguish a system from its environment by computing the closedness of the system \citep{maturana1991autopoiesis, rosen1991life, pattee2012evolving, luhmann1995probleme}. Closedness itself can be further quantified by information theory.

\begin{figure}
					\includegraphics[width=\textwidth]{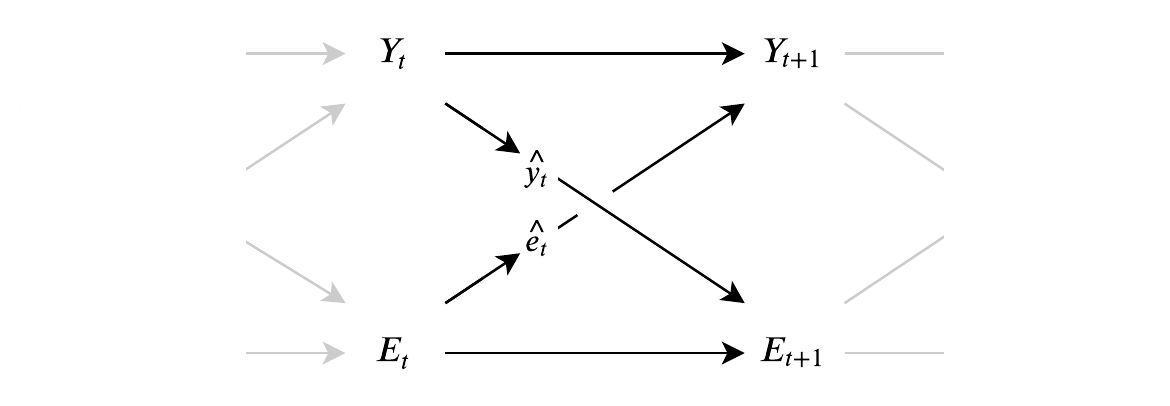}
					\caption{Dependencies between a system  $Y$ and its environment $E$ through the channels $\hat{y_t}$ and $\hat{e_t}$.} % Figure adapted from \cite{BERTSCHINGER.2006}
					\label{fig:SystemAndEnv}
				\end{figure}

		% Definition of informational closure
			Consider two processes, the environment process $(E_t)_{t \in \mathbb{N}}$ and the system process $(Y_t)_{t \in \mathbb{N}}$ and let their interaction be described by the Bayesian network with the sensor channel $\hat{e}_{t}$ and the action $\hat{y}_{t}$ channel in Fig.~\ref{fig:SystemAndEnv}. Information flow $J_{t}$ from the environment $E$ to a system $S$ at time $t$ can then be defined as the conditional mutual information $I$ between the current environment state $E_{t}$  and the future system state $Y_{t+1}$ given the current system state $Y_{t}$

				\begin{equation}
    				\label{eq:InformationFlow}
    				\left.\begin{array}
    				{rl}{J_{t}(E \rightarrow Y )} & {:= I(Y_{t+1};E_{t}|Y_{t})} \\
    				{ } & { \ = I(Y_{t+1};E_{t}) - (I(Y_{t+1};Y_{t})-I(Y_{t+1};Y_{t}|E_{t}))}
    				\end{array}\right.
				\end{equation}

			\noindent
			\cite{BERTSCHINGER.2006} defines a system as informationally closed when information flow from the environment to the system is zero.

				\begin{equation}
				J_{t}(E \rightarrow Y )=0
				\label{eq:informationflow2}
				\end{equation}

			% ------------------------------- trivial case ------------------------------- %
			\noindent
			Information closure (minimising $J_t$) is trivial if the environment and the system are entirely independent of each other.

				\begin{equation}
				\begin{aligned}
				{I(Y_{t+1};E_{t})=0}&&{\Rightarrow}&&{J_{t}(E \rightarrow Y )=0}
				\end{aligned}
				\end{equation}

			% -------------------------------- non-trivial ------------------------------- %
			\noindent
			However, informational closure can be formed non-trivially. In the non-trivial case, even though a system contains (or encodes) information about the environmental dynamics, the system can still be informationally closed. In such cases, the mutual information between the current states of the environment and the future state of the system is larger than zero.

				\begin{equation}
				I(Y_{t+1};E_{t}) > 0
				\end{equation}

			\noindent
			This also implies
				\begin{equation}
					I(Y_{t+1};Y_{t})-I(Y_{t+1};Y_{t}|E_{t}) > 0
				\end{equation}

			\noindent
			And, non-trivial informational closure can be defined as
				\begin{align}
%				\label{eq:NTIC}
    				% \left.\begin{array}
    				NTIC_t(E\rightarrow Y) :&=I(Y_{t+1};Y_{t})-I(Y_{t+1};Y_{t}|E_{t}) 
    				\label{eq:NTIC1}\\
    				&=I(Y_{t+1};E_{t})-I(Y_{t+1};E_{t}|Y_{t}) 
    				\label{eq:NTIC2}
    				% \end{array} \right.
				\end{align}

			\noindent
			Hence, maximising $NTIC_t(E\rightarrow Y)$ amounts to
				\begin{equation}
    				\label{eq:nticObjective}
    				\begin{aligned}
    				& \text{maximising} & { } & I(Y_{t+1};Y_{t}) & { } & \text{and} \\
    				& \text{minimising} & { } & I(Y_{t+1};Y_{t}|E_{t}) & { }
    				\end{aligned}
				\end{equation}
			
			\noindent
			One can also maximise $NTIC_t(E\rightarrow Y)$ by 
				\begin{equation}
    				\label{eq:nticObjective2}
    				\begin{aligned}
    				& \text{maximising} & { } & I(Y_{t+1};E_{t}) & { } & \text{and} \\
    				& \text{minimising} & { } & I(Y_{t+1};E_{t}|Y_{t}) & { }
    				\end{aligned}
				\end{equation}			

			\noindent
			This implies that the system contains within itself all the information about its own future and the self-predictive information contains the information about the environment.
			
			For simplicity, in what follows, we refer to \textit{NTIC processes} as those \textit{processes with positive NTIC}.
			
			\subsection{Informational Closure Does not Imply Causality}\label{sec:causality}
                A surprising result from the definition of information flow $J_t(E\rightarrow Y)$ (Eq.~\ref{eq:InformationFlow}) is that information flow does not indicate causal dependency from $E_t$ to $Y_{t+1}$ or from $Y_t$ to $Y_{t+1}$. Here we consider two scenarios, \textit{modelling} and \textit{passive adaptation}, which were previously noted by \citet{BERTSCHINGER.2006}. In both scenarios, a process can form positive NTIC ($NTIC(E\rightarrow Y)> 0$) and informational closure ($J(E\rightarrow Y)=0$), albeit via different causal dependencies.
                
                In the \textit{modelling} scenario, to achieve positive NTIC and informational closure, a system can internalise and synchronise with the dynamics of the environment, e.g., model the environment. In this case, the future internal state $Y_{t+1}$ of the system is driven by the current internal state $Y_t$ and the system still retains mutual information with the environment. Having high degrees of NTIC then entails high predictive power about the environment. This gives biological agents functional and evolutionary advantages.
                
                In the \textit{passive adaptation} scenario, the future system states ($Y_{t+1}$) are entirely driven by the current environment states ($E_t$). The system, perhaps counterintuitively, can nonetheless achieve positive NTIC and informational closure. This happens under the condition that the sensory process $\hat{e}_t$ is deterministic and the system merely copies the sensory values. The system is then a copy of another informationally closed process ($\hat{e}_t$) and is therefore closed. At the same time, the system has mutual information with the process that it is copying. 
                
                In most of the realistic cases, however, the environment is partially observable from the system's perspective, and thus the sensory process is usually not deterministic. Accordingly, it is difficult for the system to be informationally closed and have higher NTIC.
                More importantly, we argue in the Appendix that whenever the environment has itself more predictable dynamics than the observations, it is possible exists for a process to achieve higher NTIC by modelling the environment than by copying the observations. 
                
                We will see that both scenarios are relevant to ICT in the following sections.

% ============================================================================ %
%                            Neural coarse-graining                            %
% ============================================================================ %
	\section{Coarse-graining in the Neural System} \label{sec:Neural coarse-graining}

		%% stochasticity at microscopic levels
		The formation of NTIC with a highly stochastic process is challenging. NTIC requires the predictability of the system state and is therefore impeded by noise in the system. Information processing at the microscopic scale (cellular scale) in neural systems suffers from multiple environmental noise sources such as sensor, cellular, electrical, and synaptic noises. For example, neurons exhibit large trial-to-trial variability at the cellular scale, and are subject to thermal fluctuations and other physical noises \citep{faisal2008noise}.

        %% Our argument
		Nevertheless, it is possible that neural systems form NTIC at certain macroscopic scales through coarse-graining of microscopic neural states. Coarse-graining refers to many-to-one or one-to-one maps which aggregate microscopic states to a macroscopic state. In other words, a number of different micro-states correspond to the same value of the macro-variable \citep{price2007causation}. Coarse-grainings can therefore form more stable and deterministic state transitions and more often form NTIC processes. For neural systems this means that a microscopically noisy neural system may still give rise to an NTIC process on a more macroscopic scale.
		
		Indeed, empirical evidence suggests that coarse-graining is a common coding strategy of the neural system by which it establishes robustness against noise at microscopic scales. For instance, the inter-spike intervals of an individual neuron are stochastic. This implies that the state of an individual neuron does not represent stable information. However, the firing rate, i.e. the average spike counts over a given time interval, is more stable and robust against noise such as the variability in inter-spike intervals. Using this temporal coarse-graining strategy, known as rate coding \citep{adrian1926impulses, gerstner2002spiking, maass2001pulsed, panzeri2015neural, stein2005neuronal}, neurons can encode stimulus intensity by increasing or decreasing their firing rate \citep{kandel2000principles}. \citep{stein2005neuronal}. The robustness of rate coding is a direct consequence of the many-to-one mapping (i.e., coarse-graining).
		
		% population code
		Population coding is another example of encoding information through coarse-graining in neural systems. In this coding scheme, information is encoded by the activation patterns of a set of neurons (a neuron population). In the population coding scheme, many states of a neuron population map to the same state of macroscopic variables which encode particular informational contents, thereby reducing the influence of noise in individual neurons. That is, stable representations can be formed through coarse-graining the high dimensional state space of a neuron population to a lower dimensional macroscopic state space \citep{kristan1997population, pouget2000information, binder2009encyclopedia, QuianQuiroga2009}. Therefore, individual neuron states (microscopic scale) are not sufficiently informative about the complete encoded contents at the population scale (macroscopic scale). Instead, coarse-grained variables are better substrates for stably encoding information and allow the neural system to ignore noisy interactions at the fine-grained scale \citep{Woodward2007-WOOCWA}.
		
        These two examples show that the known coding schemes can be viewed as coarse-graining, and provide stochastic neural systems with the ability to form more stable and deterministic macroscopic processes for encoding and processing information reliably. We argue that coarse-graining allows neural systems to form NTIC processes at macroscopic scales. Based on the merit of coarse-graining in neural systems, we propose a new theory of consciousness in the next section.

% ============================================================================ %
%               A neural coarse graining theory of consciousness               %
% ============================================================================ %
	\section{\acl{OurTheory}}\label{sec:OurTheory}
	
        In this section, we propose a new theoretical framework of consciousness: the \acf{OurTheory}. The main hypothesis is that conscious processes are captured by what we call \textit{C-processes}. We first define C-processes, then state our hypothesis and discuss its implications.

		\begin{figure}[H]
		    \centering
			\includegraphics[width=\textwidth]{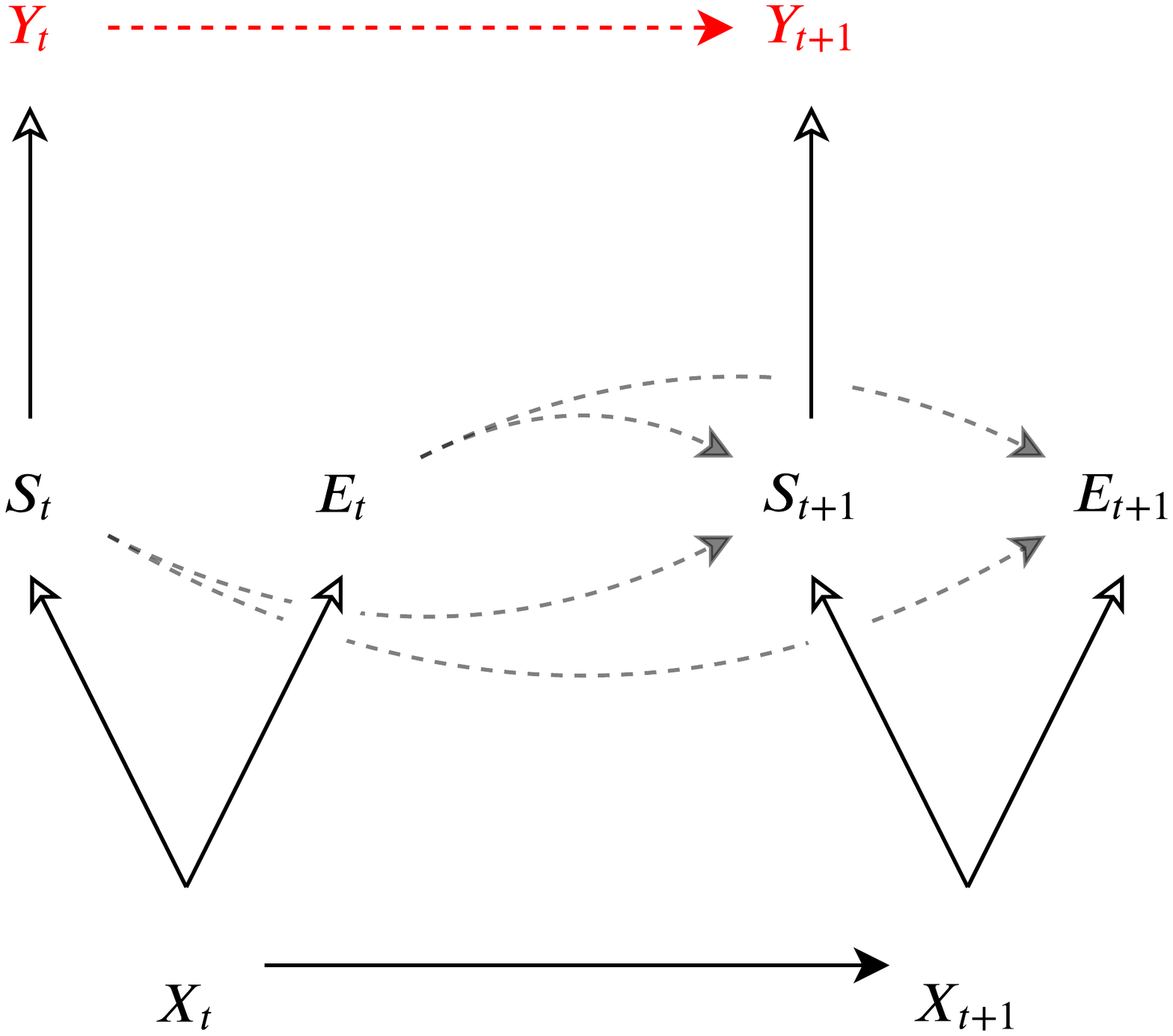}
			\caption{The information flow amounts the universe $X$, the system $S$, the environment of the system $E$, and the coarse-grained process $Y$ of the system $S$. The solid line with a filled arrow from $X_t$ to $X_{t+1}$ represents the microscopic dynamic of the universe. The solid lines with a empty arrow represent directions of coarse-graining. The dashed lines represents virtual dependencies between  two macroscopic variables. The red $Y_t$, $Y_{t+1}$, and the red dashed line in between represents a macroscopic process which forms informational closure at a certain coarse-grained scale.}
			\label{fig:fullgraph}
	   	\end{figure}

	   	% ------------------------------- definition 1 ------------------------------- %
	   	To define C-processes we first need to define coarse-grainings. Every coarse-graining is characterised by a function that maps the microscopic process to the coarse-grained macroscopic process. More formally: 
	   	
	   	\begin{definition}
	   	 Given a stochastic process $X$ with state space $\mathcal{X}$, a \emph{coarse-graining of $X$} is a stochastic process $Y$ with state space $\mathcal{Y}$ such that there exists a function \footnote{Functions in the mathematical sense used here are always either one-to-one or many-to-one.} $f_Y:\mathcal{X}\rightarrow \mathcal{Y}$ with $Y_t=f_Y(X_t)$.  
	   	\end{definition}
	   	A more general definition of coarse-grainings that maps temporally extended sequences of the microscopic process to macroscopic states are possible, but for this first exposure of our theory the simpler definition above is sufficient. 
	   	
	   	% ------------------------------- definition 2 ------------------------------- %
        \begin{definition}
        Given a stochastic process $X$ called the universe process, a \emph{C-process} is a coarse-graining $Y$ of $X$ such that the following two conditions are satisfied (see Fig.~\ref{fig:fullgraph}):
        \begin{enumerate}
        \item $Y$ is informationally closed to $X$
        \item there exists a pair $(S,E)$ of coarse-grainings of $X$ such that 
        \begin{itemize}
            \item $Y$ is a coarse-graining of $S$,
            \item the state space $\mathcal{X}$ of $X$ is equal to the Cartesian product of the state spaces $\mathcal{S}$ and $\mathcal{E}$ of processes $S$ and $E$ respectively, formally $\mathcal{X}=\mathcal{S}\times\mathcal{E}$, and 
            \item $Y$ is NTIC to $E$, formally:
        \begin{align}
            NTIC_t(E\rightarrow Y) >0
        \end{align}
        \end{itemize} 
        \end{enumerate}
       \end{definition}
       
    % ==================================== %
    % martin.biehl: 
    % doesn't work because the information closrue with respect to $X$ implies full closrure with respect to every 
    % other process. This means all conscious processes are fully closed. This contradicts what we say below.
    % Apr 7, 2020 3:23 AM
    % ==================================== %
    %     \begin{definition}
    %     Given a stochastic process $X$ called the universe process, a \emph{C-process} is a coarse-graining $Y$ of $X$ such that the following two conditions are satisfied (see Fig.~\ref{fig:fullgraph}):
    %     \begin{enumerate}
    %     \item $Y$ is informationally closed to $X$
    %     \item there exists a pair $(S,E)$ of coarse-grainings of $X$ called the \textit{system} $S$ and its \textit{environment} $E$  such that 
    %     \begin{itemize}
    %         \item $Y$ is a coarse-graining of $S$,
    %         \item the state space $\mathcal{X}$ of $X$ is equal to the Cartesian product of the state spaces $\mathcal{S}$ and $\mathcal{E}$ of processes $S$ and $E$ respectively, formally $\mathcal{X}=\mathcal{S}\times\mathcal{E}$, and 
    %         \item $E$ is the environment process of $S$ with respect to which $Y$ is maximally NTIC:
    %         \begin{align}
    %         f_E:=\argmax_{\{\E||\E|\leq |\X|\},g:\X \rightarrow \E} NTIC(g(X)\rightarrow Y) \end{align}
    %     %     $Y$ is NTIC to $E$, formally:
    %     % \begin{align}
    %     %     NTIC_t(E\rightarrow Y) >0
    %     % \end{align}
    %     \end{itemize} 
    %     \end{enumerate}
    %   \end{definition}
    % ====================================%
                               
        \noindent Note that, here we applied the same definitions of information flow (Eq.~\ref{eq:InformationFlow})
        \begin{align}
            J_{t}(E \rightarrow Y )=I(Y_{t+1};E_t|Y_t)
        \end{align}
            to the system-environment dependency
            and the micro-macro scale dependency
        \begin{align}
            J_{t}(X \rightarrow Y )=I(Y_{t+1};X_t|Y_t)
        \end{align}
        even though the Bayesian graphs differ in the two scenarios. Both these settings have been previously used in the literature \citep[see][]{BERTSCHINGER.2006, PFANTE.2014}.\\
        
         With the two definitions we can state the main hypothesis of \ac{OurTheory}
        \begin{hypothesis}
        A process $Y$ is conscious if and only if it is a C-process of some process $X$. Also the content of consciousness $C_t^{Content}$ at time $t$ is the state $y_t$ of the C-process at time $t$ and the level of consciousness $C_t^{Level}$ is the degree of NTIC of the process to the environment i.e. $NTIC_t(E\rightarrow Y)$:
        \begin{align}
            C_t^{Content} &= y_t \label{eq:cContent}\\
            C_t^{Level} &= NTIC_t(E\rightarrow Y) \label{eq:cLevel}
        \end{align}
        \end{hypothesis}

        A concrete example in the context of neuroscience is that $X$ represents the microscopic scale of the universe, $S$ a cellular scale process in the neural system, $Y$ a more macroscopic process of the neural system coarse-grained from the cellular scale process $S$, and $E$ the environment which the cellular level process $S$ interacts with.   The environment $E$ may include other processes in the neural system, the sensors for perception and interoception, and external physical worlds.
        
        Based on the hypothesis, \ac{OurTheory} leads to five core implications: 
        \begin{description}
            \item[Implication 1.] 
            Consciousness is information. Here, "informative" refers to the resolution of uncertainty. Being in a certain conscious state rules out other possible conscious states. Therefore, every conscious percept resolves some amount of uncertainty and provides information. \\ 
            This implication is also in agreement with the "axiom" of \textit{information} in Integrated Information Theory (IIT 3.0) which claims that \textquote[{\citealp[P.~2]{oizumi2014phenomenology}}]{\dots an experience of pure darkness is what it is by differing, in its particular way, from an immense number of other possible experiences\dots}
            
            \item[Implication 2.] 
            Consciousness is associated with physical substrates and the self-information of the conscious percept is equal to the self-information of the corresponding physical event. This is a direct implication from our hypothesis that every conscious percept $C_t^{Content}$ corresponds to a physical event $y_t$.

            \item[Implication 3.] 
            Conscious processes are self-determining. This is a direct implication of the requirement that $Y$ is informationally closed with respect to $X$. To be informationally closed with respect to $X$, no coarse-graining knows anything about the conscious process' future that the conscious process does not know itself. This self-determining characteristics is also consistent with our daily life conscious experience which often shows stability and continuity and is ignorant of the stochasticity (e.g., noise) of the cellular scales.
            
            \item[Implication 4.] 
            Conscious processes encode the environmental influence on itself. This is due to the non-triviality of the informational closure of $Y$ to $E$. At the same time all of this information is known to the conscious processes themselves since they are informationally closed with respect to their environments. This also suggests that conscious processes can model the environmental influence without knowing more information from the environment.
            
            \item[Implication 5.]
            Conscious processes can model environmental information (by forming NTIC) but be ignorant to part of the information of more microscopic processes (from Implication 3 and 4). This is consistent with our conscious experience, namely that the information that every conscious percept provides represents rich and structured environmental states without involving all the information about microscopic activities.
        \end{description}

		% ---------------------------------------------------------------------------- %
        %        Level of Consciousness correlates degrees of NTIC of a process        %
        % ---------------------------------------------------------------------------- %
	    \subsection{Level of Consciousness is Equal to the Degree of NTIC of a C-process}\label{sec:cl}
            According to Eq.~\ref{eq:nticObjective}, \ac{OurTheory} implies that conscious levels are determined by two quantities. 
            
            First, to form a high level of NTIC, one can increase the mutual information $I(Y_{t+1};Y_{t})$ between the current internal state $Y_t$ and the future internal state $Y_{t+1}$. In other words, conscious levels are associated with the degree of self-predictive information \citep{bialek2001predictability}. This mutual information term can be further decomposed to two information entropy quantities: 
            
            \begin{equation}
            \label{eq:SelfEntropy}
            I(Y_{t+1};Y_{t}) = H(Y_{t+1}) - H(Y_{t+1}|Y_t)
            \end{equation}
            
            This implies that a highly NTIC process must have rich dynamics with self-predictability over time. Another implication is that complex systems can potentially attain higher levels of consciousness due to the greater information capacities needed to attain high mutual information. This outcome is consistent with the common intuition that conscious levels are often associated with the degree of complexity of a system.
    
    	    Second, one can minimise the conditional mutual information $I(Y_{t+1};Y_{t}|E_{t})$ to increase the level of NTIC. If the mutual information term $I(Y_{t+1};Y_t)$ is supposed to stay large, this quantity suggests that conscious level increases with the amount of information about the environment state $E_t$ that the NTIC process encodes in its own state $Y_t$ and $Y_{t+1}$. In other words, $Y_t$ should not contain more information about $Y_{t+1}$ than $E_t$. An important implication is that agents interacting with a complex environment have the chance to build a higher level of NTIC within their systems than those living in a simple environment. In other words, the level of consciousness is associated with environmental complexity. 
    	   
    	    %% not monotonic
    		It is important to note that NTIC can be a non-monotonic function of the scale of coarse-graining. Since we can quantify the scale of a coarse-grained variable by the size of its state space, therefore, at the finest scale we consider the whole universe $X$ as the process $Y$. Then, since $Y$ is a coarse-graining of $S$ we have $Y=S=X$. In this case the environment $E$ corresponding to the universe seen as a system is the constant coarse-graining\footnote{Recall that, for a system with state space $\S$ the environment state space $\E$ must be such that $\X=\S \times \E$. If $\S =\X$ then we need $\E$ with $\X \times \E=\X$ such that $\E$ must be a singleton set. All coarse-grainings mapping $\X$ to a singleton set are constant over $\X$.} and therefore the mutual information $I(E_t;Y_{t+1})$ and the transfer entropy $I(Y_{t+1};E_t|Y_t)$ are zero. The NTIC of the universe with respect to its environment is then zero, and $X$ can never be a C-process.
    		
    		If we now increase the scale of $Y$, this allows $S$ to also reduce in scale and therefore $E$ can become more and more fine-grained. This means that the mutual information $I(E_t;Y_{t+1})$ between $E$ and $Y$ can at least potentially become positive. Up to the point where $E$ accounts for half of the bits of $X$ and $S$ for the other half the upper bound of the mutual information $I(E_t;Y_{t+1})$ achieved when $Y=S$ increases. Refining $E$ even further again leads to a reduction of the upper bound of $I(E_t;Y_{t+1})$.
            % ==================================== %
            %    WHY DO WE NEED the FUCKING $Y$    %
            % ==================================== %
            % 2020/04/07 01:57
    		% Why does Definition 2 have a fucking $Y$? It seems the whole thing could be defined wihtout referring to the underlying $S$ and directly use $Y$'s own environment (together with which it corresponds to $X$). We would always have a refinement of $Y$ that could be called $S$ to gether with its environment. But this is alwyas the case so it doesn't restrict the definition of $Y$ at all. (Especially if we can just choose $S=Y$). The only thing that makes the current definition of $Y$ more general is that it may be that $Y$ is only closed with respect to a coarsening of its own environment and not to its own environment itself. This is a waekening of the restriction compared to requiring that $Y$ be NTIC with respect to its own environment and IC with respect to $X$. Actually no, since $Y$ is IC with respect $X$ it is IC with respect to everything including to any coarsening of its own environment. Since the mutual information between $Y$ and its environment is greater or equal to the mutual information between $Y$ and a coarsening of its environment $Y$ must also be NTIC with respect to its own environment. 
    		% ==================================== %
    		
\begin{figure}[H]				
        		\includegraphics[width=\textwidth]{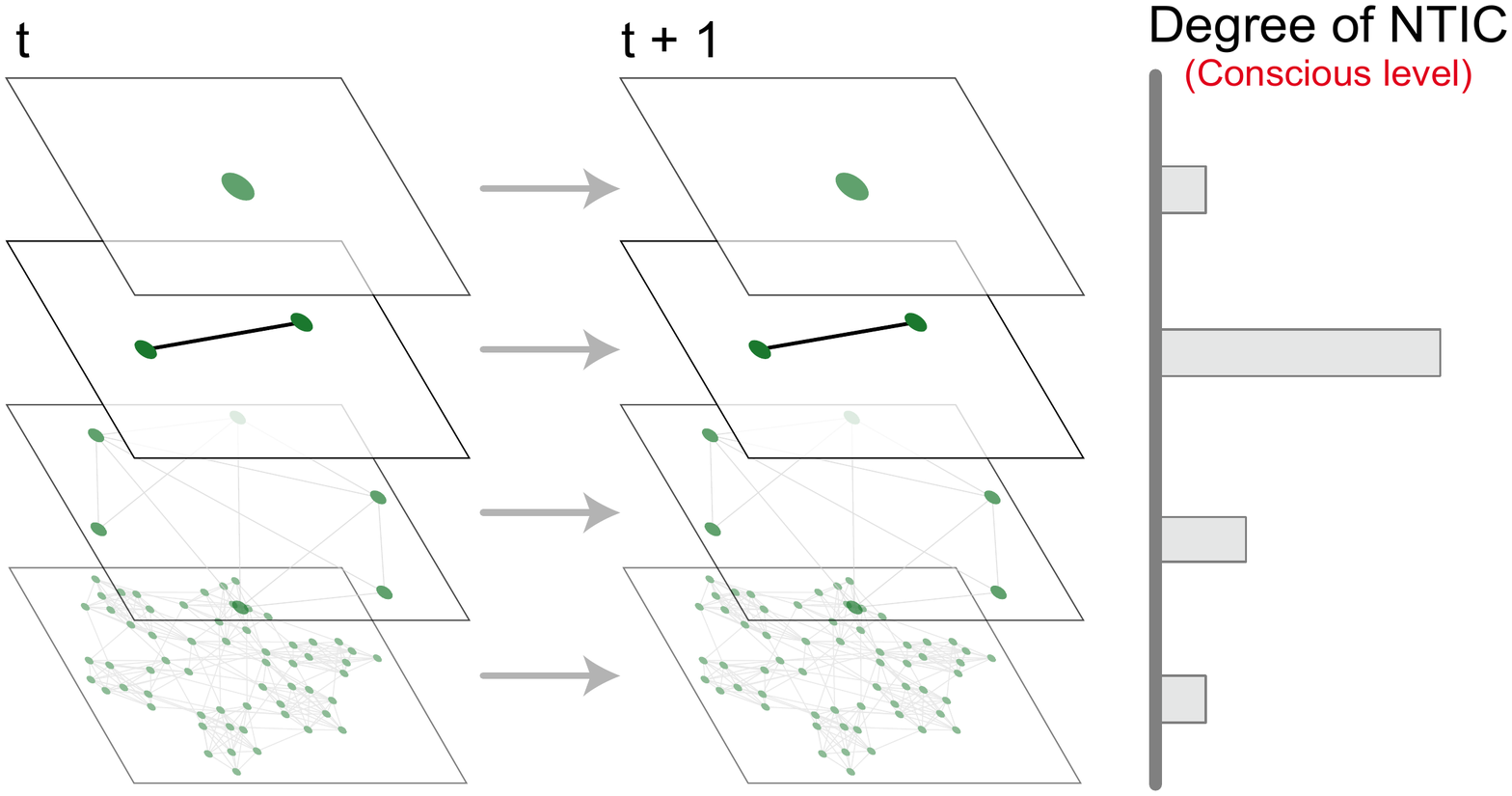}
        		\caption{A non-monotonic relationship between the scale of coarse-graining and level of consciousness.}
        		\label{fig:LevelOfConsciousness}
    		\end{figure}

    		At the other extreme, when $E=X$ the system state space must be the singleton set and NTIC from $E$ to $Y$ must again be zero. Therefore, processes at intermediate scales of coarse-graining can form higher degrees of NTIC than those at the most microscopic or macroscopic scales (Fig.~\ref{fig:LevelOfConsciousness}). \ac{OurTheory} suggests that human consciousness occurs at a scale of coarse-graining where high NTIC is formed within the neural system\footnote{In our current setup, the size of the state space $\mathcal{S}$ and $\mathcal{E}$ correspondingly determines the scale of coarse-graining of $S$ and $E$.Further research is needed to reveal the relationship among NTIC, scales of coarse-graining, and different constructions of $S$ and $E$.}. 
    		
		\subsection{Conscious Contents Corresponding to States of a C-Process}\label{sec:cc}
    	    % richness 
    		\ac{OurTheory} proposes that conscious contents correspond to the states of C-processes (Eq.~\ref{eq:cContent}). This implies that the size of the state space of a C-process is associated with the richness of the conscious contents that the process can potentially have. Accordingly, a complex C-process with a high dimensional state space can have richer conscious experience than a simple C-process. This outcome is consistent with the intuition that the richness of conscious contents is associated with the complexity of a system. 
    		
    		% CC doesn't include all microscopic states
    		Informational closure can happen between scales of coarse-graining within a single system. Thus, a macroscopic NTIC process can be ignorant of its microscopic states. \ac{OurTheory} argues that human conscious contents do not reflect cellular scale activity because the conscious process which corresponds to a macroscopic NTIC process is informationally closed to the cellular scale in the human neural system. Further more, since C-processes are informationally closed, each of them can be considered as a reality. When the information flow from its microscopic processes (and from the environment) to it is zero (Eq.~\ref{eq:informationflow2}), the future states of the process can be entirely self-determined by its past states. 
    		
    		Importantly, in most realistic cases, NTIC processes internalise the environmental dynamics in its states (see Sec. \ref{sec:causality} and also \cite{BERTSCHINGER.2006}). This suggests that an NTIC process can be considered as a process that models the environmental dynamics. This implication fits well with several theories of consciousness (for example, world simulation metaphor ~\citep{revonsuo2006inner}). Note that \ac{OurTheory} does not assume that generative models are necessary for consciousness. The implication is a natural result of processes with NTIC. 
            
            Finally, a coarse-graining can be a many-to-one map from microscopic to macroscopic states and \ac{OurTheory} proposes that conscious contents $C^{Content}$ is the state of the C-process. \ac{OurTheory} therefore implies the multiple realisation thesis of consciousness \citep{putnam1967psychological,bechtel1999multiple}, which suggests that different physical implementations could map to the same conscious experience.

		% ---------------------------------------------------------------------------- %
        %             Reconciling the levels and contents of consciousness             %
        % ---------------------------------------------------------------------------- %
	    \subsection{Reconciling the Levels and Contents of Consciousness}\label{sec:reconcile}
    	    While it is useful to distinguish the levels and contents of consciousness at the notion level, whether they can be clearly dissociated has been a matter of debate \citep{bayne2016there, Fazekas2016}. In \ac{OurTheory}, conscious levels and conscious contents are simply two different properties of NTIC processes, and the two aspects of consciousness are therefore naturally reconciled. In an NTIC process with a large state space, conscious contents should also consist of rich and high dimensional information. This framework therefore integrates the levels and the contents of consciousness in a coherent fashion by providing explicit formal definitions of the two notions.

    	    According to Sec.~\ref{sec:cl} and Sec.~\ref{sec:cc}, an important implication from \ac{OurTheory} is that both conscious levels and conscious contents are associated with the state space of an NTIC process $Y$. A larger state space of $Y$ contributes conscious levels through the mutual information $I(Y_{t+1};Y_{t})$ and also contributes richer conscious contents by providing a greater number of possible states of conscious processes. 
    	    \ac{OurTheory} therefore explains why, in normal physiological states, conscious levels and conscious contents are often positively correlated \citep{laureys2005neural}. This implication is also consistent with the intuition that consciousness is often associated with complex systems.
    
    % ============================================================================ %
    %                    Conscious versus Unconscious Processing                   %
    % ============================================================================ %
	\section{Conscious Versus Unconscious Processing}\label{sec:Conscious versus Unconscious Processing}
	    In this section, we show how \ac{OurTheory} can explain and make predictions about which processes are more conscious than others. \ac{OurTheory} is constructed using information theory and can provide predictions based on mathematical definitions. 
		
        % ---------------------------------------------------------------------------- %
        %                            Unconscious Processing                            %
        % ---------------------------------------------------------------------------- %
        \subsection{Unconscious Processing}
            In this section we highlight two scenarios in which ICT predicts that processes remain unconscious. 
            % e or  decrease the degree of NTIC of a process, thereby making the process less conscious or unconscious.
        
            \subsubsection*{Processes that are not Informationally Closed}\label{sec:reflexive}
                % ------------------------------------
                % This section should follow:
                % 1. Processes that are not informationally closed is unconscious
                % 2. Some informationally not closed processes are passively driven by the environment (some examples)
                % 3. This may give readers the impression that processes passively driven by the environment are all not informationally closed
                % 4. This is a wrong impression because in a specific circumstance which the sensory processes are deterministic, passive adaptation can be informational closed.
                % 5. However, 4 is difficult to get because in realistic cases, the environment is often partially observable and this leads to sensory processes being non-deterministic. Therefore, ICT predict that in most of the realistic cases, passive adaptation (e.g., feed-forward network, video camera) is unconscious.
                % ------------------------------------
                 
                % To maximise NTIC, one can minimise the information flow $J_{t}(E \rightarrow Y )$ from the environment $E$ to the process $Y$ (Eq.~\ref{eq:InformationFlow},\ref{eq:nticObjective2}). When the information flow is zero, the process is informationally closed with respect to the environment. 
                The first scenario is built upon the assumption that sensor processes are non-deterministic \footnote{Non-deterministic sensor processes here means $H(\hat{e}_{t+1}|\hat{e}_t)>0$.} and that process dynamics are passively driven by  environmental inputs. Such processes cannot be informationally closed and are, therefore, unconscious.
                
\begin{figure}[H]
        			\includegraphics[width=0.8\textwidth]{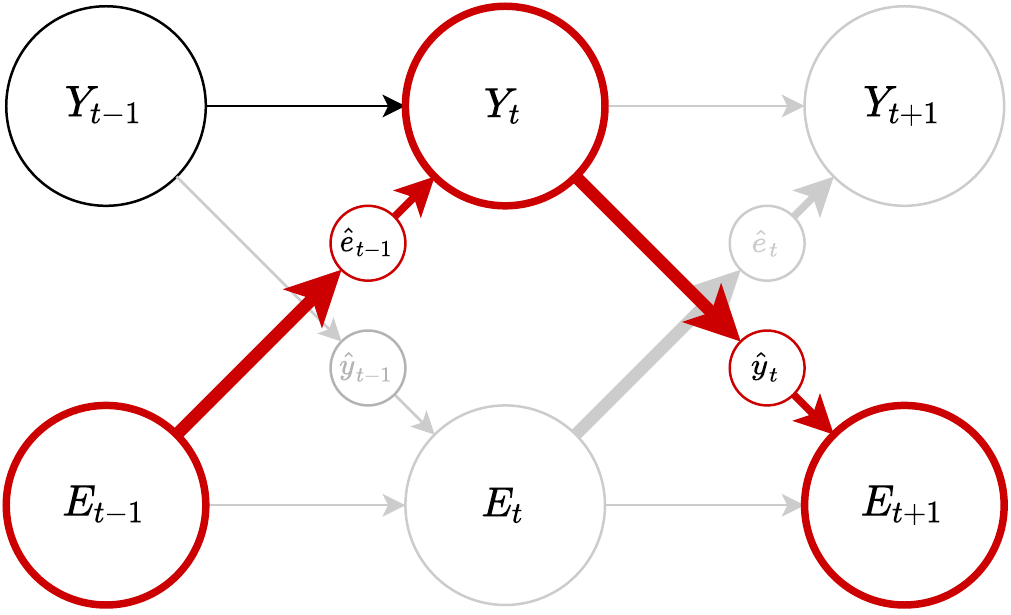}
        			\caption{
        			    Schema depicting the information flow in reflexive behaviours (shown by the red nodes and arrows) happening through the interaction between a process $Y$ and its environment $E$. When the sensor process $\hat{e}_t$ is  non-deterministic and the internal state $Y_t$ is mostly dependent on the sensor state $\hat{e}_t$ driven by the environment $E_{t-1}$ but less on its past state $Y_{t-1}$, as a consequence, $Y$ is unable to form informational closure and, therefore, remain unconscious.}
        			\label{fig:reflexive}
        		\end{figure}

                Reflexive behaviours \citep{casali2013theoretically} can be considered an example of this scenario. In \ac{OurTheory}, we can view reflexive behaviours as situations in which (Fig.~\ref{fig:reflexive}) the internal state $Y_t$, which triggers reflexive action $\hat{y}_t$, is determined by the environment state $E_{t-1}$, overruling the influences from its own past $Y_{t-1}$. Such interpretation of reflexive behaviour from the viewpoint of \ac{OurTheory} naturally explains why reflexes involve less or no conscious experience of external stimuli.
                
                The same principle can be applied to interpret blindsight \citep{humphrey1999history, humphrey1974vision, Humphrey1970} and procedural memory \citep{doyon2009contributions, ashby2010cortical} which are often considered unconscious processes.
        		Blindsight patients are able to track objects, avoid obstacles, and make above chance-level visual judgements with degraded or missing visual experience.(However, in some cases, they may still preserve some forms of conscious experience; See \cite{overgaard2011visual, mazzi2016blind}). We argue that blindsight-guided actions are a result of 
        		stimulus-response mapping. The corresponding neural circuits are driven passively and therefore are not informationally closed. According to \ac{OurTheory} we therefore have no conscious visual experience of visual stimuli.
        		
        		Similarly, for procedural memory, the state transitions of corresponding neural circuits determining the action sequences largely depend on  sensory inputs. This prevents the neural processes of procedural memory from informational closure and being conscious.
        		\ac{OurTheory} also offers an interpretation as to why patients with visual apperceptive agnosia \citep{james2003ventral} can perform online motor controls without visual awareness of action targets \citep{10.3389/fneur.2014.00255}.
        		
                Note that, not all processes that are driven by the environment (passive adaptation) are unconscious. As mentioned in Sec.~\ref{sec:causality}, when the sensor processes are deterministic, a system can still have positive NTIC and achieve informational closure via passive adaptation. Therefore, some passive system (for example pure feedforward networks) can potentially be conscious.\footnote{Since an $n$-layer feedforward network is a system with $n$-step memory it is technically appropriate to use the $n$-step memory definition of NTIC, i.e.~$NTIC^m_t(E\rightarrow Y):=I(Y_{t+1}:E_t,\dots,E_{t-n+1})-I(Y_{t+1}:E_t,\dots,E_{t-n+1}|Y_t)$ \citep{BERTSCHINGER.2006}, for such systems. In this case the notion of non-deterministic input processes should be generalised to input processes with $H(\hat{e}_t|\hat{e}_{t-1},\dots,\hat{e}_{t-n})>0$.}
                
                For agents such as human beings, the environment is often informationally rich but only partially observable in such a way that the current sensory inputs are insufficient to predict the next inputs and to form deterministic sensor processes. In this situation, the system cannot become informationally closed by passive adaptation (e.g., simply copying the sensory values to the system). ICT predicts that, in most realistic cases, processes with passive adaptation are unconscious. On the other hand, networks with recurrent loops employing information stored in their own past states have the potential to achieve higher NTIC by modelling the environment. If it turns out to be true that for every pure feed-forward network there are non-feed-forward systems achieving higher NTIC, then ICT predicts that the latter systems achieve higher levels of consciousness. This implication coincides with theories of consciousness emphasising the importance of recurrent circuits to consciousness \citep{lamme2006towards, edelman1992bright, tononi2008neural}.
            \subsubsection*{Processes that are Trivially Closed}
                The second scenario is that when encoded information in a process is trivial, i.e. there is no mutual information between the process states and the environment states $I(Y_{t+1};E_{t})$ (Eq.~\ref{eq:nticObjective2}), this leads to non-positive NTIC. In such cases, the process is considered to be unconscious. This implies that an isolated process which is informationally closed is insufficient to be conscious. 
                This mathematical property of \ac{OurTheory} 
                is relevant for 
                % provides a natural and intuitive (but only partial, see the current challenge in  Sec.~\ref{sec:Limitation and Future work}) solution to 
                dealing with 
                the boundary and individuality problems of consciousness\footnote{The boundary problem of consciousness refers to identifying physical boundaries of conscious processes and the individuality problem of consciousness refers to identifying individual consciousnesses in the universe.}
                \citep{Raymont2006-RAYUOC}. Consider an NTIC process $Y$ and an isolated informationally closed process $\hat{Y}$ with only trivial information. Adding $\hat{Y}$ to $Y$ can still maintain informational closure but does not increase non-trivial information, i.e.,  consciousness is unaffected. 
                
    			\begin{equation}
    			    \begin{aligned}
                        I(Y,\hat{Y};E) & = H(Y,\hat{Y}) - H(Y,\hat{Y}|E) \\
                                       & = H(Y) + H(\hat{Y}|Y) - (H(Y|E)+H(\hat{Y}|Y,E)) \\
                                       & = H(Y) + H(\hat{Y}) - (H(Y|E)+H(\hat{Y})) \\
                                       & = H(Y) - H(Y|E)\\
                                       & = I(Y;E)				
    				\end{aligned}
    			\end{equation}
                
                This implies that isolated processes with trivial information do not contribute consciousness and should be considered as being outside the informational boundary of the conscious processing. 
                % (for more details of the boundary detection procedure, see \cite{krakauer2014information}). 
                This property also implies that consciousnesses do not emerge from simple aggregation of informationally closed (isolated) processes which contain trivial information.
                In the future we hope to adapt the procedures for boundary detection proposed in \citet{krakauer2014information,krakauer2020information} to ICT.
                
                %\footnote{However, we do not exclude the possibility that triviality of encoded information is just a proxy of other quantities which more directly associate with consciousness. We point out that this is a limitation of the current version of \ac{OurTheory} in Sec.~\ref{sec:Limitation and Future work}.}

        % ---------------------------------------------------------------------------- %
        %                             Conscious Processing                             %
        % ---------------------------------------------------------------------------- %
		\subsection{Conscious Processing}\label{sec:conscious processing}
		    In accordance with \ac{OurTheory}, we claim that any process, system, or cognitive function which involves any C-process should be accompanied by conscious experience. 
		    
		    Previous consciousness research has identified a number of diverse cognitive processes which are often accompanied by conscious experience. \ac{OurTheory} provides an integrated account of why these processes involve conscious experience. As mentioned above, an NTIC process can be seen as an internal modelling engine for agent-environmental interactions \citep{BERTSCHINGER.2006}. Therefore, information encoded in NTIC processes is essential for several cognitive processes. 
		    
		    Among the most valuable types of information are predictions about environmental states. Cognitive functions requiring agent-scale environmental predictions are likely to recruit NTIC processes, and to therefore be accompanied by conscious experience; examples include planning and achieving long term goals.	
		    
		    %Cognitive functions involving simulations are expected to involve NTIC processes. Consequently, mental simulation, imagination, computing alternative realities, and generating counterfactuals often come with conscious experience. 	             
            
            Second, as a modelling engine, an NTIC process with a given initial state can self-evolve and simulate the environmental transitions. Cognitive functions involving internal simulations about agent-environment interactions (e.g. imagination, computing alternative realities, and generating counterfactuals) are expected to involve NTIC processes. We speculate that, these internal simulations may involve interactions between C-processes and other processes in the neural system. Therefore, they often come with conscious experience.
            
    	    Third, as an informationally closed system, an NTIC process can still provide environmental information without new sensory inputs. This is crucial for many types of off-line processing. Therefore, in contrast to reflexive-like behaviours, such as those mentioned above (Sec.~\ref{sec:reflexive}), behaviours requiring off-line computations \citep{milner1999paradoxical, himmelbach2005dorsal,revol2003pointing} often involve conscious experience. 
    	    
    	    Finally, for agents adapting to complex environments (e.g., human beings), any state of the NTIC process can be seen as an integration of high dimensional information. To accurately encode information about complex environmental states and transitions, the NTIC process requires knowledge about the complex causal dependencies involved in the environment. Cognitive functions requiring larger scale integration are therefore likely to involve C-processes and accompanied by conscious experience. % Furthermore, the high dimensional integrated information is also critical for an agent to learn new behaviours to achieve flexibility and the strategic behavioural controls and decision-making. 
    	    
    	    Note that many of the claims above are compatible with several theories of consciousness which highlight the connection between consciousness and internal simulation, predictive mechanism, or generative models inside a system (e.g. world simulation metaphor ~\citep{revonsuo2006inner}, predictive processing and Bayesian brain~\citep{clark_2013,Hohwy2013,seth2014predictive}, generative model and information generation~\citep{kanai_chang_yu_de_abril_biehl_guttenberg_2019}). Instead of relating functional or mechanistic aspects of a system to consciousness, \ac{OurTheory} captures common informational properties underlying those cognitive functions which are associated with consciousness. As such, \ac{OurTheory} does not assume any functionalist perspectives of consciousness, which associate specific functions to consciousness.  That is to say, since \ac{OurTheory} associates information  with consciousness, functional features accompanied by consciousness are collateral consequences of neural systems which utilise NTIC processes for adaptive functions. 
    	    
    	    In sum, we argue that cognitive functions involving the C-process are inevitably accompanied by consciousness. Having an NTIC process is potentially an effective approach to increasing fitness in the evolutionary process. It is likely that biological creatures evolve NTIC processes at some point during their evolution. Due to the fundamental relation between information and consciousness, biological creatures also evolve different degrees of consciousness depending on the physical scale and complexity of the environments they adapt to. 
    	    
    	    Although it starts with a non-functional hypothesis, \ac{OurTheory} accounts for the association between function and consciousness. Further, \ac{OurTheory}  demonstrates remarkable explanatory power for various findings concerning conscious and unconscious processing. 
	    
    % % ============================================================================ %
    % %                        Comparison with other theories                        %
    % % ============================================================================ %
    \section{Comparison with Other Relevant Theories of Consciousness}\label{sec:Comparison with other theories}
    In this section, we compare \ac{OurTheory} with other relevant theories of consciousness.

        % ---------------------------------------------------------------------------- %
        %        Theories about multilevel views on consciousness and cognition        %
        % ---------------------------------------------------------------------------- %
        \subsection{Multilevel Views on Consciousness and Cognition}\label{sec:MultiLevelView}
    		\ac{OurTheory} proposes that conscious processes can occur at any scale of coarse-graining which forms NTIC within a system. This suggests that the scale of coarse-graining is critical for in searching for and identifying the information corresponding to consciousness. A few number of versions of multilevel views on consciousness have previously been (explicitly or implicitly) proposed. To our knowledge, Pennartz’s neurorepresentational theory (also called Neurorepresentationalism,~\citep{pennartz2018consciousness,pennartz2015brain}) is closest to the multilevel view of \ac{OurTheory}. Similar to Neurorepresentationalism, the concept of levels in \ac{OurTheory} is relevant to Marr's level of analysis~\citep{marr1982vision, pennartz2015brain, pennartz2018consciousness}. 
    		However, ICT suggests that coarse-graining is necessary only when a process is not informationally closed. Therefore, if a C-process is formed at a microscopic scale (e.g. the scale of individual neurons), according to ICT, this C-process is sufficient for consciousness.
    		Another fundamental difference between \ac{OurTheory} and Neurorepresentationalism is that Neurorepresentationalism takes a functionalist perspective and suggests that consciousness should serve high-level world-modelling and make a best guess about the interaction between the body and the environment. 
    		%Neurorepresentationalism also suggests conscious experience is associated with integrated representation for multimodal and situational information.
    		In contrast, however, \ac{OurTheory} is grounded by a non-functional informational hypothesis. Therefore, \ac{OurTheory} provides a non-functional and fundamental explanation for the scale problem of consciousness. 
    		
    		Another well-known proposal based on multilevel views is the Intermediate Level Theory of Consciousness \citep[ILT]{prinz2007intermediate, jackendoff1987consciousness}. ILT proposes that conscious experience is only associated with neural representations at intermediate \textbf{levels of the sensory processing hierarchy} (e.g., the 2.5D representation of visual processing), and not with lower (e.g., pixel) or higher (e.g., abstract) levels of the sensory hierarchy. 
    	
\begin{figure}[H]
    			\includegraphics[width=\textwidth]{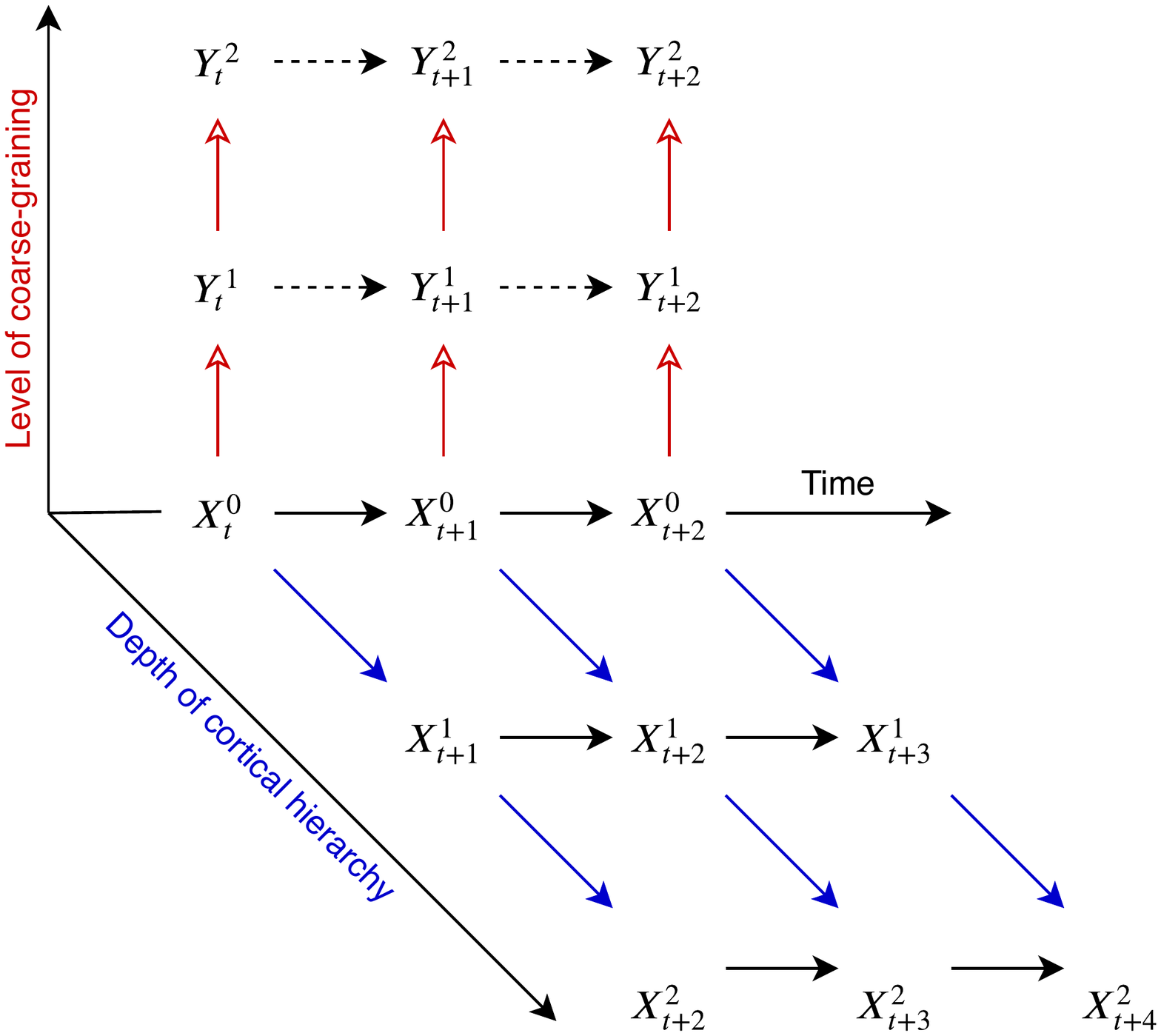}
				\caption{Distinction between the level of coarse-graining and the level of cortical hierarchy. $X$ and $Y$ represent the microscopic and macroscopic coarse-grained  variables, respectively. $X^0$ represents microscopic states upstream of the cortical hierarchy. The red empty arrows represents the directions of coarse-graining and the blue arrows represent the directions of the physical dependencies in the cortical hierarchy from upstream to downstream. (Some variables and dependencies are omitted for clarity.)}
				\label{fig:hierarchy}
    		\end{figure}

    		Here, we want to make clear that "level" in \ac{OurTheory} refers to the \textbf{scale of coarse-graining}, rather than "level" in cortical anatomy or sensory processing. It is important to note that the coarse-graining direction is an orthogonal dimension irrespective of the level of anatomy or of information processing hierarchy in the neural system (see Fig.~\ref{fig:hierarchy}). Because ILT focuses the levels of the sensory processing hierarchy and \ac{OurTheory} focus on informational closure among the levels of coarse-graining, the two theories are fundamentally different.

        % ---------------------------------------------------------------------------- %
        %                                      IIT                                     %
        % ---------------------------------------------------------------------------- %
		\subsection{Integrated Information Theory}
            Integrated information theory (IIT) states that consciousness is integrated information and that a system’s consciousness is determined by its causal properties \citep{tononi2016integrated}. \ac{OurTheory} is consistent with IIT in that informational properties are thought to underlie consciousness. In this section, we will discuss \ac{OurTheory} in the light of IIT.            
            
            \textbf{The concept of "information"}: In IIT, information refers to "integrated information", namely \textquote[\citealt{tononi2016integrated}]{Information that is specified by a system that is irreducible to that specified by its parts.} In \ac{OurTheory}, information refers to "self-information", i.e. information about the states of conscious experience and the physical states of a process. Therefore, IIT focuses more on the relationships between consciousness and causal interactions among elements within a system, whereas \ac{OurTheory} focuses more on the informational relationships between conscious experience and being in a certain state of a process. 
		    
		    \textbf{The "Exclusion" axiom in IIT}: In IIT, the Exclusion axiom claims that among all overlapping sets of elements, only one set, having maximal integrated information, can be conscious. The exclusion axiom should be applied over elements, space, time, and scales \citep{oizumi2014phenomenology, hoel2016can}.  Differing from IIT, ICT allows multiple consciousnesses to coexist across different scales of coarse-graining within a system if they are informationally closed from to each other. The two distinctive predictions decisively pinpoint the core concepts of the two theories. 

		    \textbf{The concept of "integration"}: In IIT, integrated information is a core concept in defining conscious individuals. In the present paper, we do not include the notion of integrated information within ICT. However, this represents one of the current weaknesses of ICT, namely that it in some cases it lacks the ability to individuate NTIC processes (i.e., the problem of individuality). We discuss this weaknesses in Sec.~\ref{sec:Limitation and Future work}.

		    \textbf{Prediction after system damage}: Prediction after system damage: ICT and IIT lead to different predictions when a system suffers from damage. Consider for example a densely connected network whose dynamics forms a C-process. If we cut the network in half, IIT predicts that this would result in two consciousnesses because elements in both networks still maintain high degrees of interaction. In contrast, ICT would predict that this operation might completely destroy informational closure of the network, and thereby render both parts unconscious. Nevertheless, this prediction is relatively premature. In the future, rigorous modelling studies will allow systematic comparisons between model predictions.

        % ---------------------------------------------------------------------------- %
        %                            Predictive Processing                             %
        % ---------------------------------------------------------------------------- %
		\subsection{Predictive Processing}
    		Predictive processing (PP) is a powerful framework which integrates several ideas from neuroscience. This emerging theoretical framework posits that neural systems constantly generate predictions about incoming sensory signals and update predictions based on prediction errors between predictions and sensory signals. According to PP, neural systems constantly perform unconscious statistical inference about hidden causes in the external environment. The perceptual contents are the "best guess" about those environment states which include these hidden causes \citep{clark_2013, Hohwy2013}. PP is well integrated with Bayesian brain hypothesis and has been used to interpret conscious perception in many domains \citep{Hohwy2013, seth2014predictive}.
    		
    		%% The gaps between pp and consciousness
    		PP is a powerful explanatory framework for diverse brain functions. However, to serve as a theory of consciousness, PP is still incomplete due to two explanatory gaps. First, the neural system is equipped with multiple predictive mechanisms, but it appears that not all of these predictive mechanisms are involved in conscious processes (e.g. mismatch negativity, \cite{naatanen2007mismatch}). PP needs to explain the difference between conscious and unconscious predictive mechanisms. 
    		
    		Second, PP can be considered as a sophisticated computation for perceptual inference. It takes von Helmholtz's conception of perception as unconscious inference. Thus, only the most probable outcome computed by the inference processes can be conscious, while other details of the computation remain unconscious. PP also needs to explain how unconscious inferences are able to give rise to conscious results. In short, while PP is often discussed in the context of consciousness, these explanatory gaps prevent PP from being a theory of consciousness. 
    		
    		% compatible things
    		ICT is well compatible with PP. Crucially, ICT further provides natural and fundamental explanations to fill the two explanatory gaps which hamper PP. According to the definition of NTIC, a process with high NTIC can be regarded as a powerful predictive machine which has accurate self-predictive information ($I(Y_{t+1};Y_{t})$, E.q.~\ref{eq:NTIC1}) and concurrently incorporates environmental information into its dynamic ($I(Y_{t+1};Y_{t}|E_{t})$, E.q.~\ref{eq:NTIC1}). This predictive nature of NTIC processes is in agreement with the core notion of PP in which the conscious contents are always the predicted (inferred) outcome of our predictive mechanisms. Second, due to the informational closure to the environment, the encoded information about its environment in an NTIC process can appear to be as "the best guess" about the external environment in the context of Bayesian inference. 
    		
    		% fill the gaps
    		Finally, therefore, why is some predictive information conscious and some are not? \ac{OurTheory} predicts that only the predictions generated from mechanisms involving the NTIC process are conscious. Note that it is not necessary for predictive processes to involve NTIC processes. A predictive process can make a prediction about the future state of its environment solely based on the current sensor states when the current sensor states and future sensor states have positive mutual information. However, this is not sufficient for a process to be informationally closed and, therefore, be conscious.
    		
            Also in accordance with ICT, we further propose that we can only be aware of the predictions of predictive processes due to informational closure to computational details of microscopic predictive processes. Acquisition by the macroscopic NTIC process is limited to the coarse-grained summary statistics of the microscopic processes. In other words, we predict that the computation of the statistical inferences of PP is implemented at microscopic (cellular) scales in the neural system. 
        
    		% finally
    		Finally, we consider that PP is a potential empirical implementation of NTIC processes. To maintain accurate information about the environment encoded in an NTIC process, one can open an information channel between the process and the environment to allow the minimal flow of information required to correct the divergence between them. This proposal is compatible with PP, which suggests that PP systems update (correct) the current estimations by computing prediction errors between predicted and real sensory inputs.

        % ---------------------------------------------------------------------------- %
        %                           Sensorimotor contingency                           %
        % ---------------------------------------------------------------------------- %
		\subsection{Sensorimotor Contingency}
			The sensorimotor contingency (SMC) theory of consciousness proposes that different types of SMCs give rise to different characteristics of conscious experience  \citep{o2001sensorimotor}. The theory radically rejects the view that conscious content is associated with the internal representations of a system. Rather, the quality of conscious experience depends on the agent’s mastery of SMCs. SMC emphasizes that the interaction between a system and its environment determines conscious experience.

    		ICT is not compatible with SMC. As mentioned in Sec.~\ref{sec:Conscious versus Unconscious Processing}, a process which directly maps the sensory states to the action states is insufficient to be NTIC. Therefore, learning contingencies between sensory inputs and action outputs do not imply NTIC. Hence, ICT predicts that having sensorimotor contingencies is neither a necessary nor a sufficient condition for consciousness. In fact, empirically, with extensive training on a sensorimotor task with a fixed contingency, the task can be gradually performed unconsciously. This indicates that strong SMCs do not contribute conscious contents. In contrast, ICT suggests that, with extensive training, the neural system establishes a neural mapping from sensory inputs to action outputs. This decreases the level of informational closure and, as a result, decrease the consciousness level of this process. This outcome better supports ICT than SMC.
    		
    		Nevertheless, ICT does appreciate the notion that interactions between a process and its environment are crucial to shaping conscious experience. As mentioned above, to form NTIC, a process needs to encode environmental transitions into its own dynamic. Therefore, information of agent-environment interaction should also be encoded in the NTIC process, and thereby shape conscious contents in a specific way.
    	    
    	    Different to classical SMC, a new version of SMC proposed by \cite{seth2014predictive, seth2015presence}, namely Predictive Processing of SensoriMotor Contingencies (PPSMC), combines SMC and the predictive processing framework together. PPSMC emphasises the important role of generative models in computing counterfactuals, inferring hidden causes of sensory signals, and linking fictive sensory signals to possible actions. According to ICT, if the generative model involves the NTIC process in the computation of counterfactuals, PPSMC will be compatible with our theory and may have strong explanatory power for some specific conscious experience.

        % ---------------------------------------------------------------------------- %
        %                            Global workspace theory                           %
        % ---------------------------------------------------------------------------- %
        
		\subsection{Global Workspace Theory}
		Global workspace theory (GWT; \cite{baars1988cognitive, baars1997theatre, baars2002conscious}) and Global Neuronal Workspace theory (GNWT; \cite{dehaene1998neuronal, dehaene2001towards, dehaene2011experimental}) state that the neural system consists of several specialised modules, and a central global workspace (GW) which integrates and broadcasts information gathered from these specialised modules. Only the information in the global workspace reaches conscious awareness, while information outside of it remains unconscious. These modules compete with each other to gain access to the GW, and the information from the winner triggers an all-or-none "ignition" in the GW. Information in the GW is broadcast to other modules. Conscious contents are then associates with the information that gains access to the internal global workspace \citep{Dehaene2017}.
		
        While GWT emphasises the importance of global information sharing as a basis of consciousness, the precise meaning of information broadcasting remains somewhat unclear if one tries to describe it more formally in the language of information theory. ICT offers one possible way to consider the meaning of broadcasting in GWT. Specifically, one could interpret the global workspace as the network of nodes wherein information is shared at the scale of NTIC and where communication is performed through macro-variables that are linked via mutual predictability. In other words, the global workspace should also be NTIC. While this link remains speculative, this interpretation encourages empirical studies into the relationship between the contents of consciousness and macrostate neural activities that are mutually predictive of each other.

% ============================================================================ %
%                                  Limitation                                  %
% ============================================================================ %
    \section{Limitations and Future Work}\label{sec:Limitation and Future work}  
        As a completely new theory of consciousness, \ac{OurTheory} is still far from completion. In the following, we discuss the current limitations and challenges of \ac{OurTheory} and point out some potential future research directions.
    
        % =============================== Theoretical  ===============================
        
        % Can't solve the hard problem
        It is important to clarify that \ac{OurTheory} does not intend to  solve the hard problems of consciousness \citep{chalmers1995facing}. Knowing the state of a conscious process does not allow us to answer "What is it like to be in this state of this process" \citep{nagel1974like}. Instead, \ac{OurTheory} focuses more on bridging consciousness and the physical world using information theory as a common language between them.
        
        % Identity problem
        The current version of \ac{OurTheory} cannot entirely solve the problem of individuality. The main issue with identifying individual consciousnesses using \ac{OurTheory} is that at the moment the environment is not uniquely defined. Once we have identified processes that are informationally closed with respect to $X$ we still have to find the environment process $E$ with respect to which we compute NTIC. However, there are usually multiple system processes $S$ of which a given $Y$ is a coarse-graining in which case there are also multiple environment processes $E$ with respect to which we could compute NTIC. 
        
        % Assuming that we have nonetheless identified two conscious processes 
        % In common cases, one can identify individual consciousnesses by finding informationally closed processes and computing the levels of NTIC of a process. This approach can also be applied to finding the boundaries of individual consciousnesses (for details of the boundary detection procedure, see \cite{krakauer2014information, krakauer2020information}). However, in some specific circumstances, individuality of consciousness is not clear. 
        A more general problem of NTIC-based individuality is that
        we can define a new process $Y$ and also its environment $E$ by recruiting two independent NTIC processes $Y^1~\&~Y^2$ and their environments $E^1~\&~E^2$, respectively. Accordingly, $Y = (Y^1,Y^2)$ and $E=(E^1,E^2)$. In such a case, the new process $Y$ will also be NTIC to $E$. The current version of \ac{OurTheory} is therefore unable to determine whether there are two smaller consciousnesses or one bigger consciousness (or for that matter 3 coexisting consciousnesses). The problem of individuality is a significant theoretical weakness of the current version of ICT. The notion of integration\footnote{Integration here refers to any high-order dependencies.} is a possible remedy for this issue, and we will address it explicitly in our future work using the concept of synergy.
        
        % About the environment
        The current version of ICT assumes that consciousness receives contribution from only non-trivial information, rather than trivial information encoded in a process. In other words, the amount of information about environmental states and dynamics encoded in a process is a key quantity for consciousness. However, we do not exclude the possibility that environmental information may simply be a proxy for other informational quantities. More theoretical work is needed to elucidate the role of environments. This issue will also be discuss in our future theoretical paper.
        
        In this article, we do not use a state-dependent formulation of NTIC. However, we believe that state-dependent NTIC is essential to describing the dynamics of conscious experience. The next version of ICT therefore requires further research using point-wise informational measures to construct state-dependent NTIC.
        
        Explaining conscious experience during dreaming is always a challenge to theories of consciousness. ICT currently does not have a specific answer to dreaming. However, we wish to emphasize that not all processes in the neural system are NTIC since some processes are not informationally closed. They mainly passively react to sensory inputs or other processes in the neural system. To the conscious (NTIC) process, the rest of the neural system and the body should also be considered as part of the environment. They retain some degree of activity during sleep and dreaming. We speculate that, during dreaming, the neural system stably forms a C-process with respect to its environment, i.e. the other parts of the neural system. At present, however, this remains mere speculation. Identification of the C-process(es) during dreaming is an important milestone in extending the scope of ICT.
        
        % ================================= Empirical ================================ 
        % finding the coarse-graining function 
        Empirically, a major challenge to \ac{OurTheory} is to find appropriate coarse-graining functions which map microscopic processes to macroscopic C-processes. This issue will become imperative in the search for neurological evidence supporting \ac{OurTheory}. Identifying such coarse-graining functions among infinite candidates \citep{price2007causation} appears to be very challenging. Nevertheless, recent theoretical and technical progress may contribute to  solving this issue. For example, the concept of \textit{causal emergence} proposed by Hoel \citep{Hoel19790, Hoel2018} has been further developed recently. Causal emergence is highly relevant to the relationship between informational closure and coarse-graining. In their new study, \cite{klein2019uncertainty}, start to compare how different coarse-graining functions influence causal emergence at macroscopic scales. \cite{PFANTE.2014, PFANTE.2014b} provide a thorough mathematical analysis of level identification, including informational closure. In neuroscience, an understanding of neural population codes has also made a tremendous progress due to advance in recording technique and data science \citep{Kohn2016, panzeri2015neural}. \cite{Gamez2016} has also systematically described relevant issues in finding data correlates of consciousness among different levels of abstraction. We believe that interdisciplinary research is required to narrow down the scope of search for coarse-graining functions and conscious processes at macro-scales in the neural system and beyond.
        
        Finally, another empirical challenge to ICT is that of empirical supporting evidence. This is understandable because the concept of NTIC is relatively new in the history of information science, not to mention in neuroscience. Very few experiments and data collections examining NTIC properties in neural systems have yet appeared. To our knowledge, only two studies \citep{Palmer2015, sederberg2018learning} coincidentally examined relevant properties in salamander retina; these found that a large group of neural populations of retinal ganglion cells encoded predictive information about external stimuli and also had high self-predictive information about their own future states. This result is consistent with the characteristic of NTIC. We expect that there will be more empirical studies examining relevant neural properties of NTIC.
        
% ============================================================================ %
%                                  Conclusion                                  %
% ============================================================================ %
    \section{Conclusions}
    % short re-introduction 
    In this paper, we introduce the \textbf{\acf{OurTheory}}, a new informational theory of consciousness. \ac{OurTheory} proposes that a process which forms informational closure with non-trivial information, i.e. \textbf{non-trivial informational closure (NTIC)} is conscious and through coarse-graining the neural system can form conscious processes, at certain macroscopic scales. \ac{OurTheory} considers that information is a common language to bridge the gap between conscious experience and physical reality. Using information theory, \ac{OurTheory} proposes computational definitions for both conscious level and conscious content. This allows \ac{OurTheory} to be generalised to any system beyond the human brain. 
    
    % For neuroscience    
    ICT provides an explanation for various findings from research into conscious and unconscious processing. The implications of ICT indicate that the scales of coarse-graining play a critical role in the search for neural substrates of consciousness. Improper measurement of neurophysiological signals, such as those which are excessively fine or coarse in scale, may lead to misleading results and misinterpretations.

    % For scientific theory of consciousness
    \ac{OurTheory} reconciles several theories of consciousness. \ac{OurTheory} indicates that they conditionally coincide with \ac{OurTheory}'s implications and predictions but, however, not the fundamental and sufficient conditions for consciousness. Example theories include those which emphasise recurrent circuits \citep{lamme2006towards, edelman1992bright}; highlight the internal simulation,  predictive mechanisms, and generative models ~\citep{revonsuo2006inner, clark_2013,Hohwy2013, kanai_chang_yu_de_abril_biehl_guttenberg_2019, seth2014predictive, seth2015presence}; and relate to multilevel view of consciousness~\citep{pennartz2018consciousness,pennartz2015brain,prinz2007intermediate, jackendoff1987consciousness}. Notably, while \ac{OurTheory} is proposed based on the non-functional hypothesis, its implications for the functional aspects of a system fit several functionalist proposals well.

	% For philosophy of mind
	Regarding philosophy of mind, \ac{OurTheory} connects several distinct arguments together. First, \ac{OurTheory} can be seen as an identity theory because it assumes a fundamental relation between consciousness and information. Second, the implications of \ac{OurTheory} tightly link consciousness to several cognitive functions in the context of evolution. This explains why people might intuitively have a functionalist point of view of consciousness. \ac{OurTheory} emphasises that informational closure between scales of coarse-graining is critical to form NTIC processes in some stochastic systems. In this case, especially for the neural system, forming conscious processes at macroscopic scales coincides with the perspective of emergentism. Finally, forming NTIC (conscious) processes through many-to-one maps, i.e., coarse-graining, implies multiple realisability of consciousness. As a result, \ac{OurTheory} provides an integrated view for these arguments and is further capable of indicating how and why they are conditionally true.
	
	% Final 
	The current version of ICT is still far from completion, and several outstanding issues mandate further theoretical and empirical research. Nevertheless, ICT offers an explanation and a prediction for consciousness science. We hope that ICT will provide a new way of thinking about and understanding of neural substrates of consciousness.  
	
	% ============================================================================ %
	%                                      End                                     %
	% ============================================================================ %
	\section*{Acknowledgements}
	\addcontentsline{toc}{section}{Acknowledgements}
	A.C., Y.Y, and R.K. are funded by the Japan Science and Technology Agency (JST) CREST project. Work by M.B. and R.K. on this publication was made possible through the support of a grant from Templeton World Charity Foundation, Inc. The opinions expressed in this publication are those of the authors and do not necessarily reflect the views of Templeton World Charity Foundation, Inc. This manuscript has been released as a Pre-Print at arXiv \citep{chang2019information}.

    \section*{Author Contributions Statement}
    \addcontentsline{toc}{section}{Author Contributions Statement}
    A.C. conceived and developed the theory. M.B. and A.C. contributed the mathematical formalisation of the theory. A.C., M.B, and R.K wrote the manuscript, based on a first draft by A.C. with extensive comments from Y.Y. All authors contributed to manuscript revision, read and approved the submitted version.

    \section*{Conflict of Interest Statement}
    \addcontentsline{toc}{section}{Conflict of Interest Statement}
    All authors were employed by Araya Inc. The authors declare that the research was conducted in the absence of any commercial or financial relationships that could be construed as a potential conflict of interest.
\section*{Appendix}
\label{sec:appendix}
            Let us assume that the system only observes a part of the environment state.
            % \footnote{In theory this part of the environment state may still evolve deterministically over time in which case the system need only copy this part to achieve NTIC and consciousness. This seems to be a rare case as well and we assume the observed part of the environment is not close to deterministic in the following.} 
            % In this case it is well known that the observed part of the environment need not even be a Markov process even if the environment itself is deterministic. So the mutual information between subsequent observations may be much lower than the mutual information between subsequent environment states.
            
            % Furthermore, in this case the transfer entropy from environment to the system becomes equal to the transfer entropy from the observed part of the environment to the system. 
            
            % Finally, note that the transfer entropy that can be achieved by copying the environment state or the observed part of the environment state equal to the mutual information between the subsequent environment states or observations. Copying the environment state corresponds to setting $S_{t+1}=E_t$ such that the transfer entropy is
            % \begin{align}
            %  I(S_{t+1}:E_t|S_t)=I(E_t:E_t|E_{t-1})=H(E_t|E_{t-1}).
            % \end{align}
            We can represent the part of the environment that we observe by the value of a function $f$ applied to the environment state. In this case we get for the transfer entropy
            \begin{align}
             I(S_{t+1}:E_t|S_t)=I(S_{t+1}:f(E_t)|S_t).
            \end{align}
            If the system only copies the observation we then get for the transfer entropy
            \begin{align}
             I(S_{t+1}:f(E_t)|S_t)=I(f(E_t):f(E_t)|f(E_{t-1}))=H(f(E_t)|f(E_{t-1}))
            \end{align}
            and for the mutual information
            \begin{align}
             I(S_{t+1};E_t)=I(f(E_{t+1});E_t)=H(f(E_t))
            \end{align}
            such that 
            \begin{align}
                NTIC_t(E\rightarrow S)=I(f(E_t);f(E_{t-1})).
            \end{align}
            This shows that whenever there is mutual information between subsequent observations a process that only copies the observations has positive NTIC. Note that any additional (internal) processing of the observation without reference to an additional internal state using a function $g$ can only reduce this mutual information:
            \begin{align}
               I(f(E_t);g(f(E_{t-1})))\leq I(f(E_t);f(E_{t-1})).
            \end{align}
            However, ignoring restrictions due to a possibly fixed choice of the universe process $X$ we find that for each such system there are other systems that achieve higher NTIC. For example, if we define the system to be the "mirrored" and synchronized environment by setting $S_t:=E_t$, then the transfer entropy vanishes
            \begin{align}
             I(S_{t+1}:E_t|S_t)=I(E_{t+1}:E_t|E_t)=0
            \end{align}
            and the mutual information is equal to the mutual information between the current and next environment state:
            \begin{align}
                I(S_{t+1};E_t)=I(E_{t+1};E_t).
            \end{align}
             In cases where the environment has itself higher predictive mutual information than the observations it produces - in other words, when
             \begin{align}                I(E_{t+1};E_t)\geq I(f(E_{t+1});f(E_t))
            \end{align}
            there is then potential for a predictive process to achieve higher NTIC than a copying system or any system that only processes its last observations without taking account of other internal memory (i.e.\ those systems also applying $g$ to their observations). Note that this also holds true in cases where the observations are themselves closed. If there is a more complex environment behind them, the mirrored and synchronised system has higher NTIC with respect to that environment than the system copying the observations.

	\bibliographystyle{authordate1}
	\addcontentsline{toc}{section}{Reference}
	\bibliography{ms}
	
	\section*{Figure Legends}
	\renewcommand{\listfigurename}{~}
	\addcontentsline{toc}{section}{Figure Legends}
    \listoffigures

\end{document}